\documentclass[a4paper,fleqn,usenatbib]{mnras}

 \usepackage[normalem]{ulem}

\usepackage[T1]{fontenc}
\usepackage{ae,aecompl}

\usepackage{graphicx}	
\usepackage{amsmath}	
\usepackage{amssymb}	

\title[HCCO in cold dense clouds]{A proposed chemical scheme for HCCO formation in cold dense clouds}

\author[Wakelam et al.]{
V. Wakelam$^{1,2}$\thanks{E-mail:
wakelam@obs.u-bordeaux1.fr}, J.-C. Loison$^{3,4}$, K. M. Hickson$^{3,4}$, M. Ruaud$^{1,2}$
\\
$^{1}$ Univ. Bordeaux, LAB, UMR 5804, F-33270, Floirac, France.\\
$^{2}$ CNRS, LAB, UMR 5804, F-33270, Floirac, France\\
$^{3}$ Univ. Bordeaux, ISM, UMR 5255, F-33400 Talence, France\\
$^4$ CNRS, ISM, UMR 5255, F-33400 Talence, France
}

\date{Accepted XXX. Received YYY; in original form ZZZ}

\pubyear{2015}

\begin{document}
\label{firstpage}
\pagerange{\pageref{firstpage}--\pageref{lastpage}}
\maketitle

\begin{abstract}
The ketenyl radical (HCCO) has recently been discovered in two cold dense clouds with a non-negligible abundance of a few $10^{-11}$ (compared to H$_2$) \citep{2015A&A...577L...5A}. Until now, no chemical network has been able to reproduce this observation. We propose here a chemical scheme that can reproduce HCCO abundances together with HCO, H$_2$CCO and CH$_3$CHO in the dark clouds Lupus-1A and L486. The main formation pathway for HCCO is the OH + CCH $\rightarrow$ HCCO + H reaction as suggested by \citet{2015A&A...577L...5A} but with a much larger rate coefficient than used in current models. Since this reaction has never been studied experimentally or theoretically, this larger value is based on a comparison with other similar systems.
\end{abstract}




\section{Introduction}

Cold dark clouds are assumed to be the most simple type of interstellar sources that can be used to test chemical models before being applied to more complex sources such as protostars and protoplanetary disks \citep{2007ARA&A..45..339B,2013ChRv..113.8710A}. The clouds have characteristically low temperatures (below 10~K), they are dense (between a few $10^4$ to a few $10^6$~cm$^{-3}$) and are shielded from any source of UV field. There exists however a source of UV photons produced by the emission from excited H$_2$ produced by electrons originating from the ioniozation of H$_2$ by cosmic-ray particles \citep{1983ApJ...267..603P}. More than 70 molecular species (neutrals, cations and anions) have now been observed in these objects \citep{2013ChRv..113.8710A}. HCCO was recently detected by \citet{2015A&A...577L...5A} with a non negligible abundance of a few $10^{-11}$ (compared to H$_2$); an abundance that they could not reproduce using their model.\\ 
To study the formation and destruction of these species, chemical models have been developed considering an increasing number of processes both in the gas-phase and at the surface of the grains where a fraction of these species are believed to be formed \citep{2009ARA&A..47..427H}. Such models are based on chemical networks containing chemical reactions with their associated rate coefficients. The networks have been constructed over time mostly driven by new observations. Over the past few years, we have been revisiting those chemical networks  from a chemists point of view, to provide a better description of the underlying processes. \citep{2009A&A...495..513W,2012MNRAS.421.1476L,2012PNAS..10910233D,2014MNRAS.437..930L,2014MNRAS.443..398L}. In this paper, we present a chemical scheme for the formation and destruction of HCCO in cold dense clouds and compare our results with the observed abundances.

\section{Chemical modeling and network}

\subsection{The NAUTILUS chemical model}\label{model}


The model used to simulate the abundance of HCCO in cold dense clouds is NAUTILUS. This model computes the abundance of molecules and atoms (neutrals but also some species in their cationic or anionic forms) in the gas-phase and also at the surface of interstellar grains. The equations and the chemical processes included in the model are described in earlier papers and we refer to \citet{2014MNRAS.440.3557R}.\\
For the gas-phase, many chemical reactions (including bimolecular and unimolecular processes) are considered based on the kida.uva.2014 chemical network by \citet{2015ApJS..217...20W}. Interstellar ice is modeled by a one-phase rate equation approximation \citep{1992ApJS...82..167H}: there is no differentiation between the species in the bulk and at the surface. The species from the gas-phase are allowed to physisorb on the surface of interstellar grains. Here, they can diffuse and react. The evaporation processes are: thermal (which are inefficient at dense cloud temperatures), induced by cosmic-rays \citep[following ][]{1993MNRAS.261...83H}, and chemical desorption \citep[as defined by ][]{2007A&A...467.1103G}. Photodesorption has not been introduced in this model because photodesorption induced by cosmic-ray secondary photons was found to be inefficient compared to the chemical desorption mechanism by \citet{2014MNRAS.445.2854W}. In addition, \citet{2012PCCP...14.9929B} showed experimentally that photodesorption of few CO layers (i.e. $\sim$10) adsorbed on amorphous  H$_2$O was inefficient. Until more experiments are performed on water ices, it is reasonable to assume that molecular photodesorption is only a minor desorption process in dense clouds. Species can diffuse by thermal hopping only with a barrier of 0.5$\times$E$_{\rm D}$ (with E$_{\rm D}$ the species binding energy). Reactions with atomic hydrogen can undergo tunneling using the formalism by \citet{1992ApJS...82..167H}. For  chemical desorption, i.e. partial desorption of the products due to the exothermicity of the reactions occurring at the surface of the grains,   approximately 1\% of the products is allowed to desorb. In addition to the usual surface diffusion reactions, we have introduced the complexation and low temperature Eley-Rideal mechanism from \citet{2015MNRAS.447.4004R}. The Cosmic-Ray Induced Diffusion mechanism is not included because it does not have any impact on the surface chemistry at high visual extinction \citep[see][]{2014MNRAS.440.3557R}. \\
To simulate the chemistry of dense clouds, the model is used with homogeneous conditions and integrated over $10^7$~yrs. The gas is initially composed of atoms (partly ionized depending on the ionization potential of the atoms). The abundances are the same as in \citet{2011A&A...530A..61H}. For fluorine, not included in Hincelin et al., we use the depleted value of $6.68\times 10^{-9}$ (compared to the total proton density) from \citet{2005ApJ...628..260N}. The C/O elemental ratio is the subject of considerable debate and is of crucial importance for the modeling of dense clouds \citep{2011A&A...530A..61H}. For our standard model, we have used a C/O ratio of 0.7 (i.e. the oxygen elemental abundance is $2.4\times 10^{-4}$). The model was run with a dust and gas temperature of 10~K, a total proton density of $2\times 10^4$~cm$^{-3}$, a cosmic-ray ionization rate of $1.3\times 10^{-17}$~s$^{-1}$, and a visual extinction of 30. 

\subsection{Modification of the network}

The initial gas-phase network is kida.uva.2014 \citet{2015ApJS..217...20W}. The network for surface reactions and gas-grain interactions is based on the one from \citet{2007A&A...467.1103G} with several additional processes from \citet{2015MNRAS.447.4004R}. The HCCO molecule was introduced by Ruaud et al. but was not discussed in their paper since it was mostly an intermediate to the formation of more complex species on the surface. Indeed, the gas-phase reactions involving this species were very limited and as we will show in the next section, the gas-phase abundance of this species was very low at times larger than $10^5$~yr. In addition to these reactions \citep[listed in Tables A.1 and A.2 of ][]{2015MNRAS.447.4004R}, we have introduced and reviewed reactions involving HCCO radicals (presented in Table~\ref{list_reactions}) to take into account the most important pathways for the formation and destruction of this species in the gas-phase.

\section{Predicted abundances}

\begin{figure}
\includegraphics[width=0.8\linewidth]{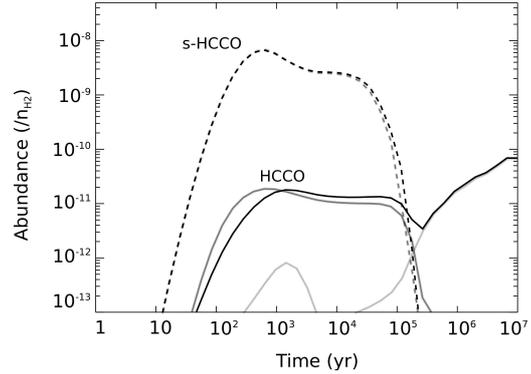}
\caption{HCCO abundance (with respect to H$_2$) predicted by our model for typical dense cloud conditions  (see section~\ref{model}) as a function of time in the gas-phase (solid lines) and at the surface of the grains (dashed lines). Grey lines show the results obtained with the previous network while the black lines show the results including the new chemistry. The light grey curve for HCCO in the gas phase has been obtained with the new chemistry but removing the reaction s-CCO + s-H  on the surface. \label{HCCO}}
\end{figure}

Using the model described above, we have run the simulations with the new chemistry and compared the predictions with the previous results. The HCCO abundance predicted by the model in the gas-phase and at the surface of the grains is shown in Fig.~\ref{HCCO}. The formation of HCCO on the surface is very efficient between $10^2$ and $10^5$~yr because of the reaction $\rm s-H + s-CCO \rightarrow s-HCCO$, which is barrier-less for the equivalent gas phase process \citep{Bauer1985,Schmidt1983,1983CPL...100..251H}. s-CCO is formed through the Eley-Rideal mechanism developed in \citet{2015MNRAS.447.4004R}: gas-phase carbon atoms react without a barrier with CO on the ice surface. The absence of a barrier for the C + s-CO $\rightarrow$ s-CCO reaction is deduced from the study of \citet{Husain1971} and ab-initio calculations performed by \citet{2015MNRAS.447.4004R}.  The gas-phase abundance of HCCO before $10^5$~yr is then the result of the chemical desorption (see section \ref{model}) of s-HCCO during its formation on the surface. The abundance of HCCO in the gas-phase predicted by the new model without this reaction is also shown in Fig.~\ref{HCCO}. Note that in Table B1 of \citet{2015MNRAS.447.4004R} only the channel leading to s-HCCO is indicated but the chemical desorption channel was also included. \\
 After $10^5$~yr, its abundance decreases in the previous model because s-HCCO is also hydrogenated to give s-H$_2$CCO and we did not have any efficient gas-phase formation processes. With the new chemistry, its abundance in the gas-phase increases at larger times. The main production reaction is $\rm OH + CCH \rightarrow HCCO + H$ as already suggested by \citet{2015A&A...577L...5A}. There are no theoretical or experimental studies of this reaction. \citeauthor{2015A&A...577L...5A} used a rate coefficient of $3\times 10^{-11}$~cm$^3$s$^{-1}$ based on \citet{Frenklach1992}. This value is however simply an estimation and there is no justification in \citet{Frenklach1992} of this choice. Based on similar systems, i.e. the O + CCH rate constant \citep{Boullart1996,1996CPL...261..450D,Devriendt1997,2011IAUS..280..372G} or OH + radicals reactions \citep{DeavillezPereira1997,Jasper2007,Sangwan2012}, we propose the use of a larger rate coefficient of $2\times 10^{-10}$~cm$^3$s$^{-1}$. This rate constant is close to the capture rate dominated at low temperature by dipole-dipole interactions and to the one precisely computed by \citet{2011IAUS..280..372G} for O + CCH. The rate constant calculation methodology has been described in detail in \citet{2014MNRAS.437..930L}.

In this case, the formation of HCCO is faster than its destruction through reactions with H, O and H$_3^+$. There is also another potential source of HCCO radicals: the CCH + O$_2$ reaction. The rate constant of this reaction is high at low temperature \citep{1998FaDi..109..165C,Vakhtin2001} and HCCO has been experimentally identified as one of the products \citep{Lange1975} with a non-negligible branching ratio between 20\% to 40\% \citep{Lange1975,Laufer1984}. The production of HCCO radicals is also suggested from theoretical calculations \citep{Sumathi1998,Li2004}. However in the \citet{Lange1975} and \citet{Laufer1984} studies, the rate constant of the CCH + O$_2$ reaction was underestimated and the rate constant of the HCCO + O$_2$ reaction was overestimated leading to a large uncertainty in the HCCO product branching ratio. Moreover our current astrochemical model leads to an O$_2$ abundance higher than observed for dense molecular clouds \citep{2003A&A...402L..77P}, inducing an overestimation of HCCO production through the CCH + O$_2$ reaction. Instead of including HCCO production from the CCH + O$_2$ reaction in our standard model \citep[the CCH + O$_2$ reaction producing only H, HCO, CO and CO$_2$ ][]{Su2000}, we performed a test run to see the effect of its inclusion. When the CCH + O$_2$ $\rightarrow$ HCCO + O reaction is included, an increase of the HCCO abundance is observed in the 3-5$\times 10^5$ yr range (reaching few $10^{-11}$ of H$_2$) corresponding to the peak of the O$_2$ abundance profile.\\
Fig.~\ref{Others} shows the abundances of HCO, H$_2$CCO and CH$_3$CHO predicted by our new model as a function of time for the physical conditions described in section~\ref{model}. Compared to \citet{2015A&A...577L...5A}, H$_2$CCO is predicted to be much less abundant very likely because we have introduced its destruction reaction with atomic carbon with a rate coefficient of $3\times 10^{-10}$~cm$^3$~s$^{-1}$ \citep[see][]{2015MNRAS.447.4004R}.  
Our predicted CH$_3$CHO abundance is quite different from the one of \citet{2015A&A...577L...5A} with the UMIST RATE12 or the kida.uva.2014 networks. 
The difference with their results using the kida.uva.2014 network is mainly due to the fact that they do not consider reactions at the surface of the grains. In our present simulation, CH$_3$CHO is produced in the gas phase through the O + C$_2$H$_5$ reaction, C$_2$H$_5$ being efficiently produced through surface reactions (mainly s-H + s-C$_2$H$_4$ $\rightarrow$ C$_2$H$_5$).  The difference between the gas phase models of \citet{2015A&A...577L...5A} using the two networks is more complex to analyse. 
One main difference though between our network and UMIST RATE12 is the inclusion of destruction reactions with atomic carbon.


\section{Comparison with observations}

\begin{figure}
\includegraphics[width=0.8\linewidth]{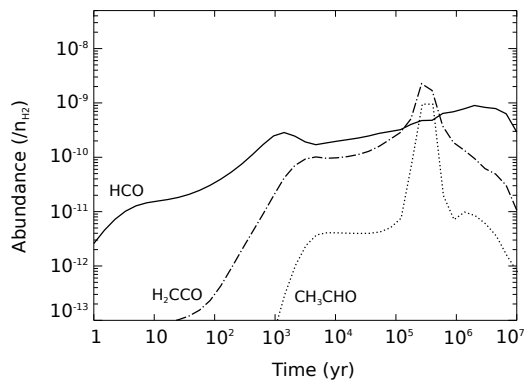}
\caption{HCO, H$_2$CCO and CH$_3$CHO abundances (with respect to H$_2$) predicted by our model using the new chemistry for the typical dense cloud conditions described in section~\ref{model} as a function of time in the gas-phase. \label{Others}}
\end{figure}

\begin{figure}
\includegraphics[width=0.8\linewidth]{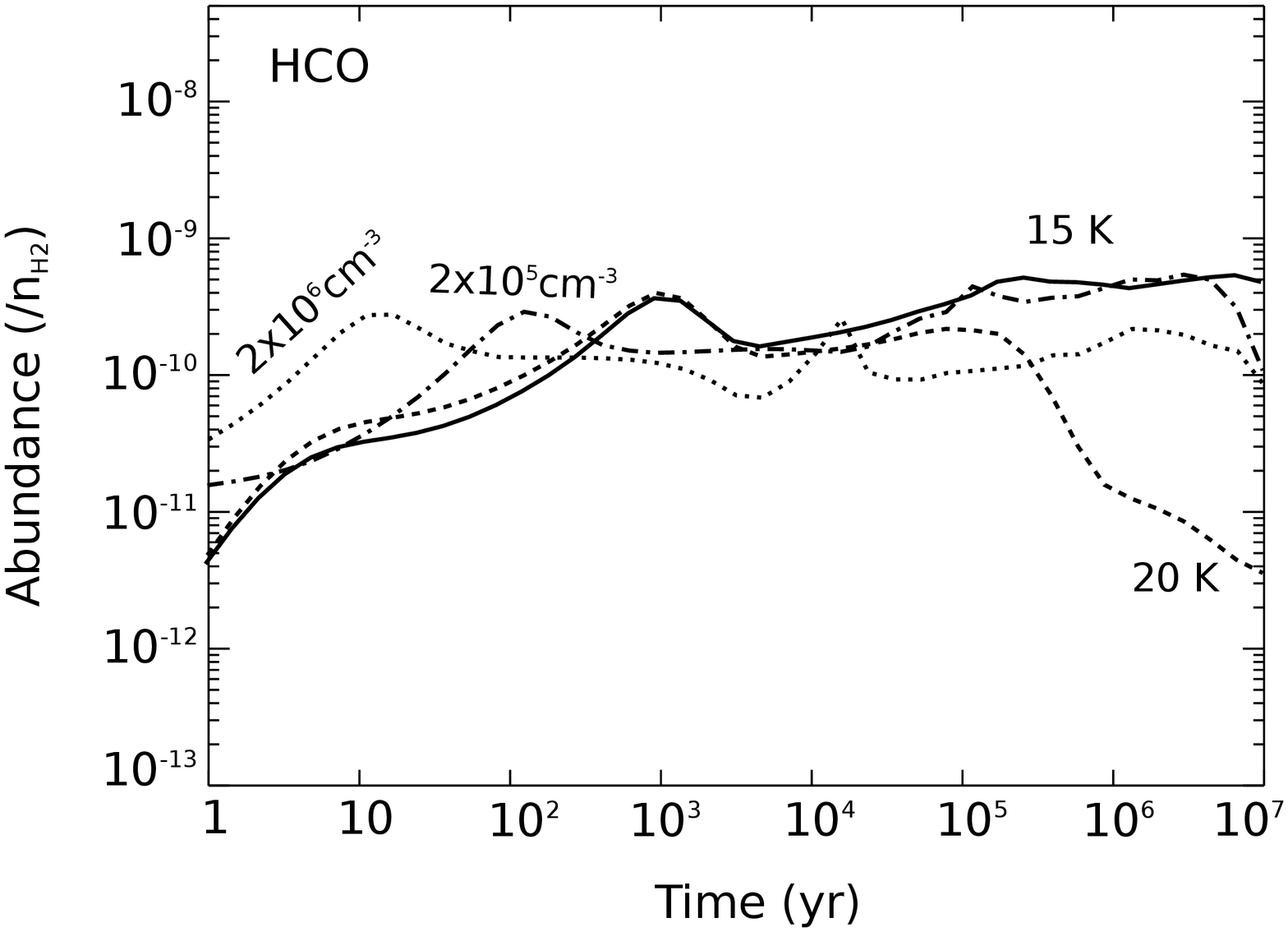}
\includegraphics[width=0.8\linewidth]{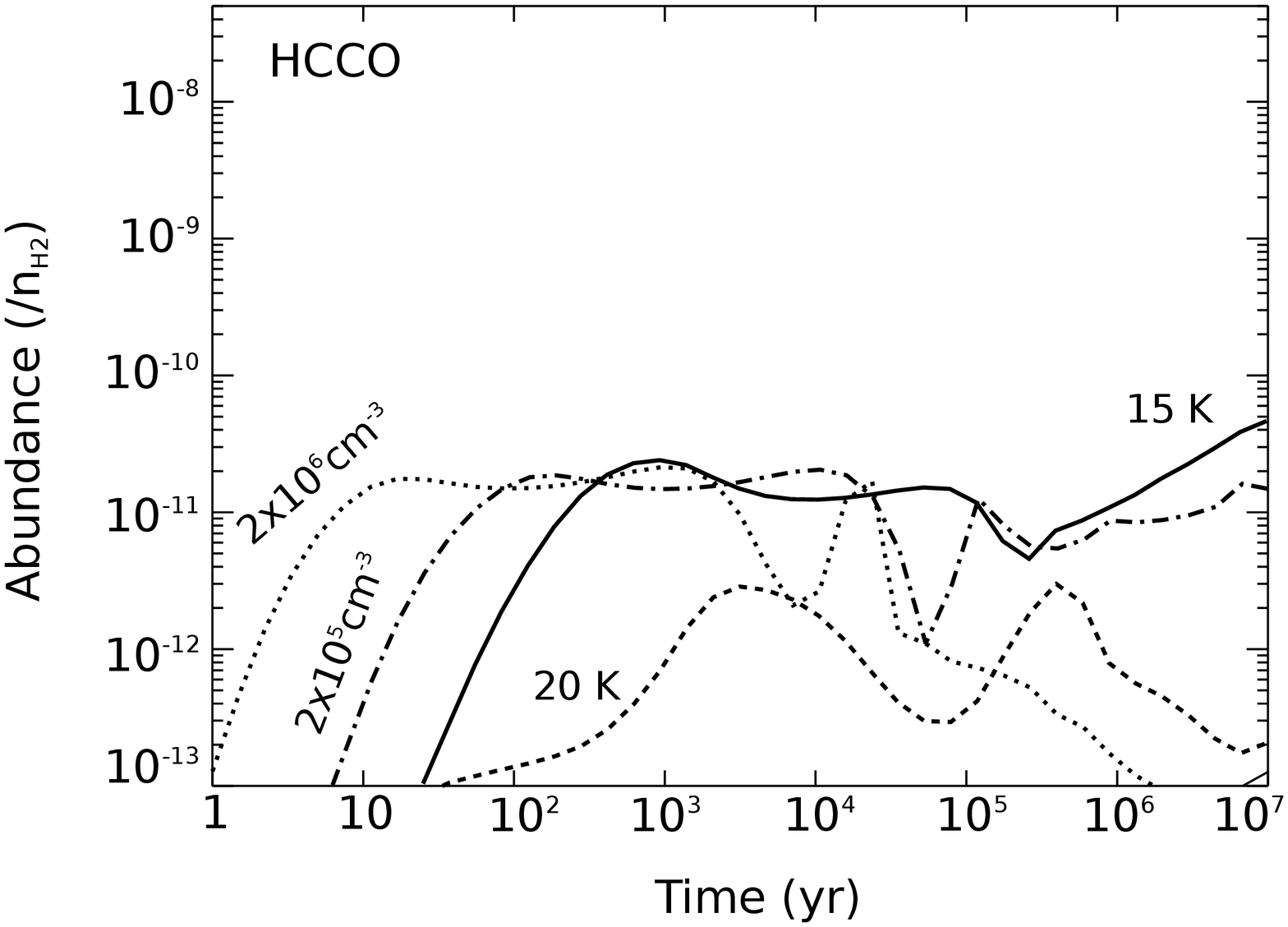}
\includegraphics[width=0.8\linewidth]{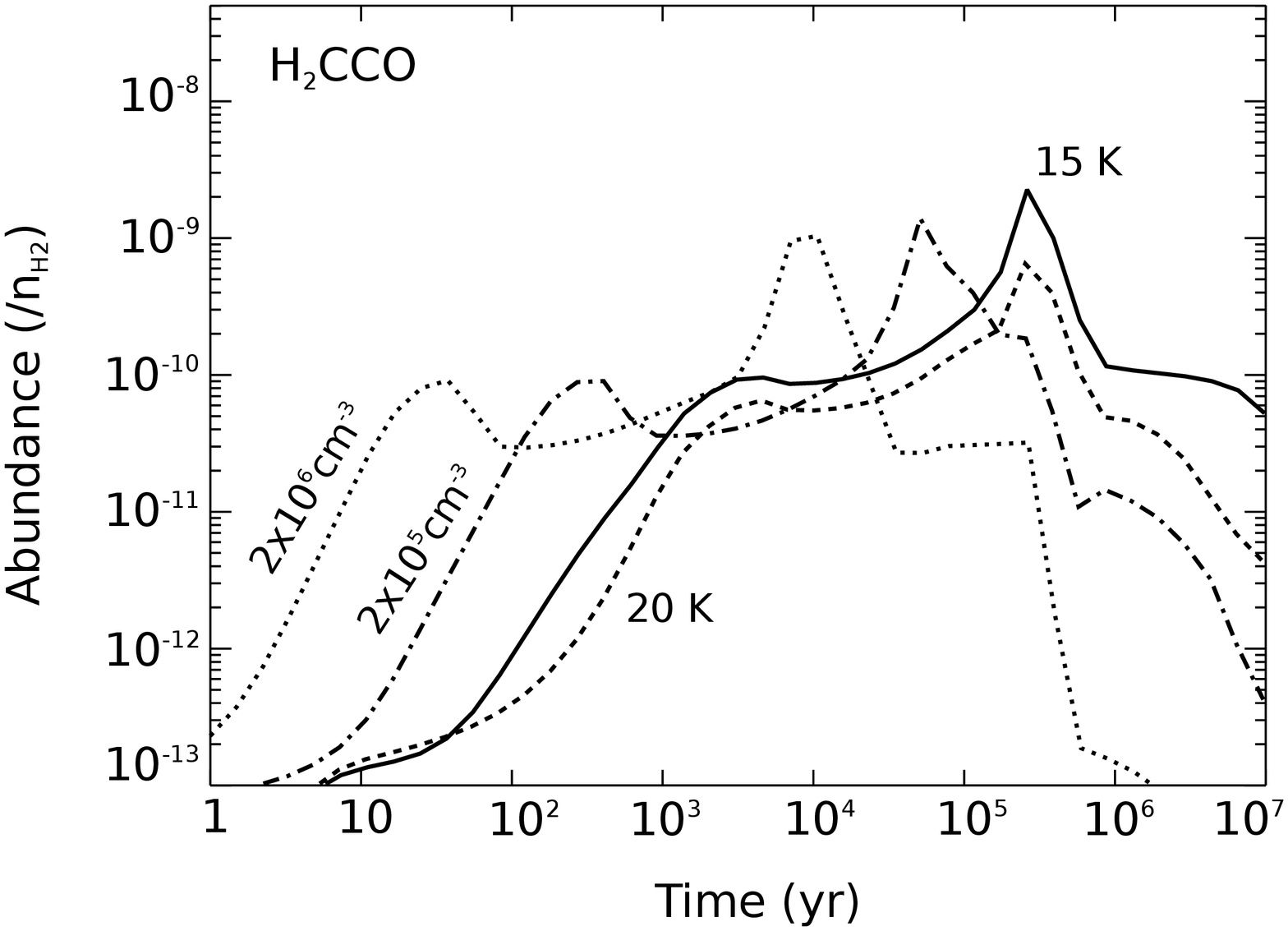}
\includegraphics[width=0.8\linewidth]{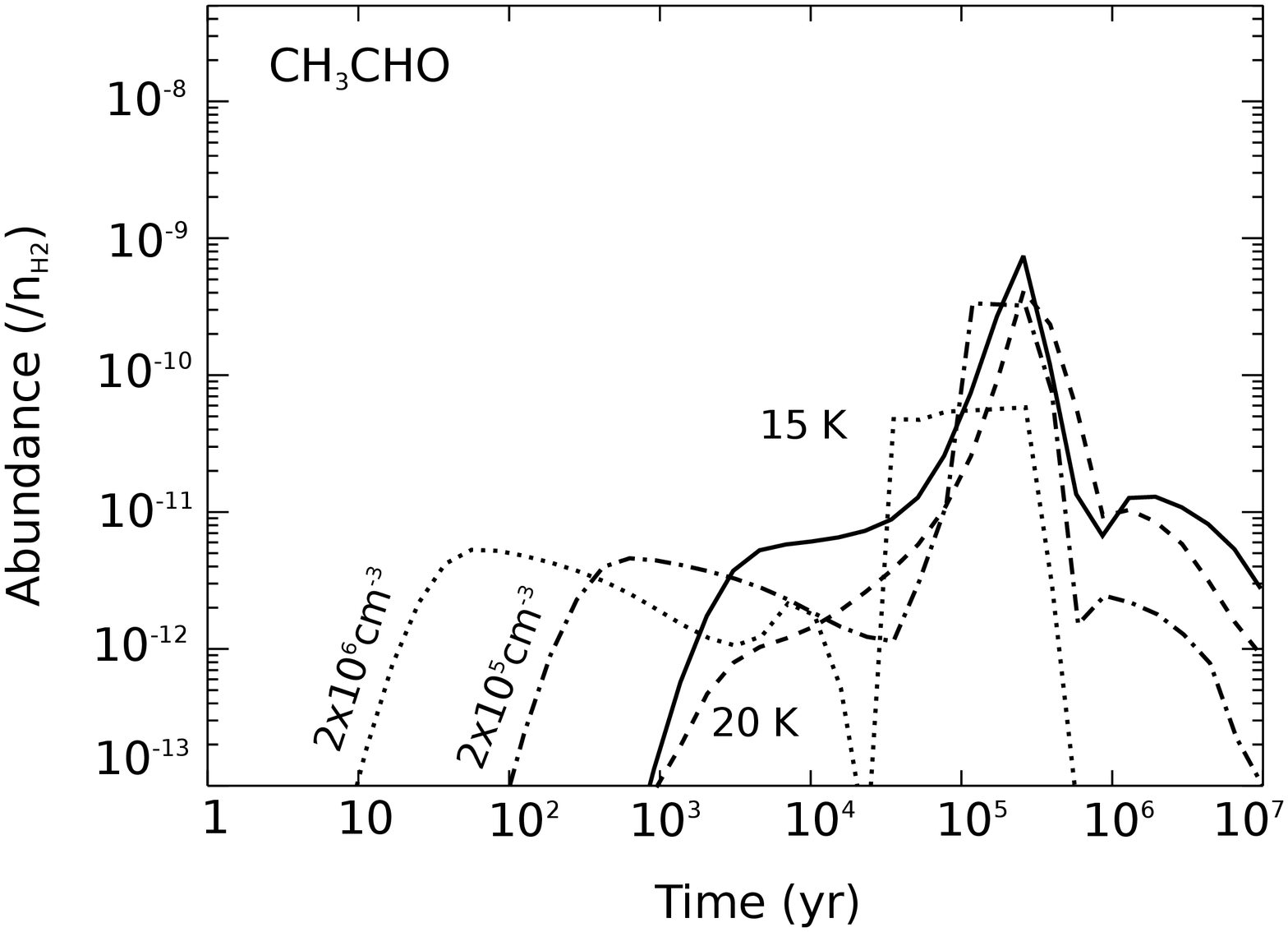}
\caption{HCO, HCCO, H$_2$CCO and CH$_3$CHO abundances (with respect to H$_2$) predicted by our model using the new chemistry as a function of time in the gas-phase. Some of the physical parameters  have been changed compared to the standard model: the gas and dust temperature is set to 15~K for the solid lines and to 20~K for the dashed lines (keeping the H density to $2\times 10^4$~cm$^{-3}$) ; the density was set to $2\times 10^5$~cm$^{-3}$ for the dotted-dashed lines and to $2\times 10^6$~cm$^{-3}$ for the dotted lines (keeping the temperature to 10~K). \label{all_models}}
\end{figure}

HCCO has only been observed in two dark clouds, Lupus-1A and L483 by \citet{2015A&A...577L...5A} with an abundance of a few $10^{-11}$ compared to H$_2$. These authors also give upper limits in three other cold sources with a very strong limit for L1521F of $7\times 10^{-13}$. They also detected HCO, H$_2$CCO and CH$_3$CHO in the same sources. We will compare our model results with the abundances listed in Table 2 of \citet{2015A&A...577L...5A} keeping in mind the uncertainties in the H$_2$ column densities in these sources \cite[see ][for discussion]{2015A&A...577L...5A}. In TMC-1, HCCO was not detected and an upper limit of $2.9\times 10^{-12}$ (compared to H$_2$) was derived by \citet{1989ApJ...340..900T}. The reanalysis of the HCCO observation by \citet{2015A&A...577L...5A} gave an upper limit of $5\times 10^{-11}$. \\
 For all the molecules (Figs.~\ref{HCCO} and \ref{Others}), we are able to reproduce reasonably well the observed abundances for times between $10^5$ and $10^6$ yr (chemical ages usually assumed for dark clouds). Only HCO is slightly overproduced by our model. It is interesting to note that the predicted HCO abundance presents a smooth increase while H$_2$CCO and in particular CH$_3$CHO have peak abundances around $2\times 10^5$~yr. This peak abundance of H$_2$CCO is ten times larger than the observed one in any source.
 The strong increase of H$_2$CCO and CH$_3$CHO near $10^5$~yr is due to the fact that these closed shell molecules react only with C atoms and not with H, O or N atoms; the carbon atom abundance strongly decreasing between $10^5$ and $2\times 10^5$~yr. The relatively stable abundance of HCO with time may explain the uniform abundance of this molecule observed by \citet{2015A&A...577L...5A} in all sources. The main production and destruction reactions for those molecules are summarized in Table~\ref{mainreactions}.\\
The observed sources present some variations in density and temperature that may influence the chemistry \citep[see][and references therein]{2015A&A...577L...5A}. To test this, we have run our model with a larger total proton density (n$_{\rm H}$ = n(H) + n(H$_2$)) of $2\times 10^5$ and even $2\times 10^6$~cm$^{-3}$ (see Fig.~\ref{all_models}), keeping the other parameters similar to the ones described in section~\ref{model}. As a general result of more collisions, the predicted gas-phase abundances are increased earlier but the peak abundances and late time abundances are smaller because of more efficient depletion on the grains. If some of the sources are denser such as Lupus-1A and L1521F, it is very likely that they have not been that dense for a long period of time. Some of the sources may also be slightly warmer than the typical 10~K usually assumed. We have run our model with gas and grain temperatures of 15 and 20~K (see Fig.~\ref{all_models}). The 15~K model is not very different from the 10~K standard model. In contrast, the 20~K model produces fewer molecules in general except for CH$_3$CHO which remains relatively unaffected. After $10^5$~yr, the 20~K regime is characterized by a smaller abundance of CCH and C$_2$H$_3$, a smaller abundance of CO onto the grain surfaces (the evaporation temperature of CO being around 18~K), a larger abundance of atomic hydrogen in the gas-phase, and an increase of the diffusion rate of heavy radicals onto the surface. Before $10^5$~yr, less HCCO is produced because of a smaller abundance of CO on the grains (less CCO is formed). After $10^5$~yr, it is the larger gas-phase abundance of atomic hydrogen that destroys HCCO efficiently in the gas-phase. The lower abundance of HCCO in the ices leads to less H$_2$CCO in the gas-phase. \\
Using an approach similar to \citet{2014MNRAS.437..930L} to compare the HCO, HCCO, H$_2$CCO and CH$_3$CHO observed and modeled abundances obtained with the standard physical model, we have found that the observed abundances in Lupus-1A and L483 are best reproduced at $1.8\times 10^5$~yr with a mean difference between observed and modeled abundances of less than a factor of 3. Using a gas and dust temperature of 15~K, the best mean agreement is smaller than a factor of 2 at $1.1\times 10^5$~yr because HCCO and H$_2$CCO are better reproduced. Considering both the uncertainties in the observations and the chemical modeling, this difference between the models is however insignificant. The "best age" that we give here is only based on the comparison between the modeled and observed abundances of the four molecules. In addition, estimating an age for such sources requires us to define the time zero as a starting point. The formation of dense clouds is a continuous process from the diffuse medium. The formation of molecules is likely to start early on during this process when the density and the visual extinction are smaller than the ones considered in our model. For these reasons, this age should only be considered as the best time in our model to reproduce the observations.


\section*{Acknowledgements}

The authors thank the following funding agencies for their partial support of this work: the French CNRS/INSU programme PCMI and the ERC Starting Grant (3DICE, grant agreement 336474).






\bibliographystyle{mnras}
\bibliography{HCCO}


\appendix

\section{List of added reactions}

\begin{table*}
\caption{List of gas-phase reactions added to the model and associated parameters.}
\begin{center}
\begin{tabular}{|l|l|l|c|c|c|c|c|c|c|}
\hline
\hline
& Reaction  & & $\Delta E$ (kJ/mol) & $\alpha$ & $\beta$ & $\gamma$ & F$_0$ & g & Reference\\
 \hline
1 & H + HCCO & $\rightarrow$ CH$_2$ + CO & -113 & $1.7\times 10^{-10}$ & 0.17 & 0 & 1.6 & 0 & 1,2 \\
\hline
2 & C$^+$ + HCCO & $\rightarrow$ C$_2$H$^+$ + CO & -411 & $2.0\times 10^{-9}$ & 0 & 0 & 3 & 0 & 3 \\
& & $\rightarrow$ HCCO$^+$ + C  & -143 & $2.0\times 10^{-9}$ & 0 & 0 & 3 & 0 &  3 \\
\hline
3 & C + HCCO & $\rightarrow$ CCH + CO & -437 & $2.0\times 10^{-10}$ & 0 & 0 & 3 & 0 & 4 \\
\hline
4 & N + HCCO & $\rightarrow$ HCN + CO & -624 & $2.0\times 10^{-11}$ & 0.17 & 0 & 3 & 10 &  5 \\
& & $\rightarrow$ HNC + CO & -569 & $1.0\times 10^{-11}$ & 0.17 & 0 & 3 & 10 &  5\\
& & \sout{$\rightarrow$ NCCO + H} & -220 & \sout{$4.0\times 10^{-11}$} & \sout{0.17} & \sout{0} & \sout{3} & \sout{10} & 5 \\
\hline
5 & O + HCCO & $\rightarrow$ H + CO + CO & -428 & $1.6\times 10^{-10}$ & 0 & 0 & 1.6 & 0 &  2\\
& & $\rightarrow$ CH + CO$_2$ & -221 & $4.9\times 10^{-11}$ & 0 & 560 & 2 & 100 &  2\\
\hline 
6 & OH + CCH & $\rightarrow$ HCCO + H & -220 & $2.0\times 10^{-10}$ & 0 & 0 & 3 & 0 &  3\\
& & $\rightarrow$ CH + CO + H & +96 & 0 & 0 & 0 & 3 & 0 &  3\\
\hline
7 & HCCO + H$_3^+$ & $\rightarrow$ H$_2$CCO$^+$ + H$_2$ & -387 & 1.0 & $3.1\times 10^{-9}$ & 2.8 &2 & 0 &  6\\
\hline
8 & HCCO + H$_3$O$^+$ & $\rightarrow$ H$_2$CCO$^+$ + H$_2$O & -122 & 1.0 & $1.4\times 10^{-9}$ & 2.8 &2 & 0 &  6\\
\hline
9 & HCCO + HCO$^+$ & $\rightarrow$ H$_2$CCO$^+$ + CO & -244 & 1.0 & $1.3\times 10^{-9}$ & 2.8 &2 & 0 &  6\\
\hline
10 & HCCO + N$_2$H$^+$ & $\rightarrow$ H$_2$CCO$^+$ + N$_2$ & -339 & 1.0 & $1.3\times 10^{-9}$ & 2.8 &2 & 0 &  6\\
\hline
11 & H$_2$CCO$^+$ + e$^-$ & $\rightarrow$ CH$_2$  + CO & -607 & $2.0\times 10^{-7}$ & -0.5 & 0 & 3 & 0 &  \\
& & $\rightarrow$ H + HCCO & -495 & $1.0\times 10^{-8}$ & -0.5 & 0 & 3 & 0 &  \\
& & $\rightarrow$ H + CH + CO & -178 & $1.0\times 10^{-7}$ & -0.5 & 0 & 3 & 0 &  \\
& & $\rightarrow$ H + H + CCO & -92 & 0 & 0 & 0 & 3 & 0 &  \\
\hline
12 & CH$_3$CO$^+$ + e$^-$ & $\rightarrow$ CH$_3$  + CO & -629 & $1.0\times 10^{-7}$ & -0.5 & 0 & 3 & 0 &  \\
& & $\rightarrow$ H + H$_2$CCO & -509 & $1.0\times 10^{-7}$ & -0.5 & 0 & 3 & 0 &  \\
& & $\rightarrow$ H + CH$_2$ + CO & -185 & $1.0\times 10^{-7}$ & -0.5 & 0 & 3 & 0 &  \\
& & $\rightarrow$ H + H + HCCO & -72 & 0 & 0 & 0 & 3 & 0 &  \\
\hline
\end{tabular}
\end{center}
$\Delta E$ is the enthalpy of reaction.\\
$\alpha, \beta, \gamma$ are the parameters to compute the temperature dependent rate coefficients. The Arrhenius-Kooij formula described in \citet{2012ApJS..199...21W} should be used for the reactions except for reactions 7 to 10 for which the ionpol1 formula from the KIDA database (http://kida.obs.u-bordeaux1.fr) should be used \citep[see also][]{2012ApJS..199...21W}. The rates can be used for temperatures between 10 and 300~K.\\
F$_0$ and g are parameters used to describe the temperature dependent uncertainty in the rate coefficient \citep[see ][]{2012ApJS..199...21W}.\\
References: \\
1 \citet{Glass2000}\\
2 \citet{2005JPCRD..34..757B}\\
3 Close to capture rate\\
4 Estimation considering the well known reactivity of carbon atoms \\
5 Reaction on triplet surface with some spin orbit crossing.  Part of the HCN should lead to HNC considering the available energy. The last channel was removed because NCCO is not in KIDA. \\
6 Capture rate \\
\label{list_reactions}
\end{table*}%

\section{Main formation and destruction pathways}

\begin{table*}
\caption{Main reactions of production and destruction for HCO, HCCO, H$_2$CCO and CH$_3$CHO for two different times and using the new model.}
\begin{center}
\begin{tabular}{|l|c|c|}
\hline
\hline
\multicolumn{3}{|c|}{$10^4$~yr} \\
Species & Formation & Destruction \\
\hline
HCO & O + CH$_2$ $\rightarrow$ H + HCO & H + HCO $\rightarrow$ CO + H$_2$ \\
 & O + C$_2$H$_4^+$ $\rightarrow$ HCO + H$_3^+$ & C + HCO $\rightarrow$ CH + CO / H + CCO \\
\hline
HCCO & s-H + s-CCO $\rightarrow$ HCCO & O + HCCO $\rightarrow$ H + CO + CO \\
  &   & C + HCCO $\rightarrow$ CCH + CO \\
\hline
H$_2$CCO & O + C$_2$H$_3$ $\rightarrow$ H$_2$CCO + H & C + H$_2$CCO $\rightarrow$ C$_2$H$_2$ + CO \\
 & s-H + s-HCCO $\rightarrow$ H$_2$CCO & \\
 \hline
 CH$_3$CHO & O + C$_2$H$_5$ $\rightarrow$ H + CH$_3$CHO & C + CH$_3$CHO $\rightarrow$ C$_2$H$_4$ + CO \\
 \hline
\multicolumn{3}{|c|}{$10^6$~yr} \\
Species & Formation & Destruction \\
 \hline
HCO & s-H + s-CO $\rightarrow$ HCO & H + HCO $\rightarrow$ CO + H$_2$\\
& H$_2$COH$^+$ + e$^-$ $\rightarrow$ H + H + HCO & \\
\hline
HCCO & OH + CCH $\rightarrow$ HCCO + H & H + HCCO $\rightarrow$ CH$_2$ + CO\\
\hline
H$_2$CCO & O + C$_2$H$_3$ $\rightarrow$ H + H$_2$CCO & H$_3^+$ + H$_2$CCO $\rightarrow$ H$_2$ + CH$_3$CHOH$^+$\\
\hline 
CH$_3$CHO & s-H + s-CH$_3$CO $\rightarrow$ CH$_3$CHO & H$_3^+$ + CH$_3$CHO $\rightarrow$ H$_2$ + CH$_3$CHOH$^+$\\
\hline
\end{tabular}
\end{center}
\label{mainreactions}
\end{table*}%



\bsp	
\label{lastpage}
\end{document}